\documentstyle[12pt]{article}
\def\be{\begin{equation}}
\def\ee{\end{equation}}
\def\beq{\begin{eqnarray}}
\def\eeq{\end{eqnarray}}

\begin{document}
\begin{titlepage}
\begin{flushright}
SU-4252-741
\\  
\end{flushright}
\begin{center}
   \vskip 3em
  {\bf \LARGE CHIRAL SUPERCONDUCTING } 
  \vskip 3em
 {\bf \LARGE  MEMBRANES }
  \vskip 3em
{\large Rub\'en Cordero${}^{(1)}$ and Efra\'{\i}n 
Rojas${}^{(2, 3)}$ \\[3em]}
\em{
${}^{(1)}$Departamento de F\'{\i}sica, \\
Escuela Superior de F\'{\i}sica y Matem\'aticas 
del IPN\\
Edificio 9, 07738, M\'exico D.F., MEXICO}\\[1em]
${}^{(2)}$ Physics Department, Syracuse University\\
Syracuse, NY 13244-1130, USA \\[1em]
${}^{(3)}$ Departamento de F\'{\i}sica \\
Centro de Investigaci\'on
y de Estudios Avanzados del I.P.N. \\
Apdo Postal 14-740, 07000 M\'exico,
D. F.,
MEXICO \\[1em]

\end{center}
\vskip 3em
\begin{abstract}
We develop the dynamics of the chiral superconducting 
membranes (with null current) in an alternative 
geometrical approach either 
making a Lagrangian description and a Hamiltonian point 
of view. Besides of this, we show the equivalence of the 
resulting descriptions to the one known Dirac-Nambu-Goto 
(DNG) case. Integrability for chiral string model is 
obtained using a proposed light-cone gauge. In a similar 
way, domain walls are integrated by means of a simple 
ansatz. We compare the results with recently works appeared 
in the literature.  
\end{abstract}


\end{titlepage}

\newpage

\section {Introduction}

It is believed that cosmic strings are fundamental 
bridges in the understanding of the Universe formation
due to that several cosmological phenomena can be 
described by means of the cosmic strings properties.
Besides of these there are other kinds of cosmic 
objects possesing different properties of those 
inherited to ordinary cosmic strings, for example:
domain walls, hybrid structures like domain walls
bounded by strings and so forth. For an extense review
in the context of cosmology, see the Ref. 
\cite{Vilenkin1}. 
They can arise in several Grand Unified Theories  
whenever there exists an appropriate symmetry breaking scheme. 
However, there is other class of cosmological objects that
can emerge with the ability to carry some sort of charge.  
For instance, as was suggested by Witten \cite{Witten} 
in the middle of the eighties, cosmic strings could 
behave like superconductors.

Since that time, the vast research on superconducting
strings has thrown a new variety of cosmic objects.
The cosmological result of supersymmetric theories
(SUSY) was also considered, yielding to another class of 
cosmic strings, namely chiral cosmic strings. These 
objects are the result of a symmetry breaking in SUSY
where a $U(1)$ symmetry is broken with a Fayet-Iliopoulos $D$ 
term, turning out a sole fermion zero mode traveling in
only one direction in the string core \cite{Davis2}.
In other words, when the current along the superconducting 
string shows a light-like causal structure then we 
have a chiral string. Carter and Peter \cite{Carter}
have made an exhaustive study of this kind of cosmic strings
and some time later on, it has been continued by other authors 
\cite{Vilenkin,Davis1,Steer,Olum}. They have found solutions
for chiral cosmic strings taking advantadge of different
gauge choices showing new cosmological properties. 
The fermionic zero mode is traveling at the speed of 
light, whereby chiral strings have a different evolution from 
DNG strings. The dynamics of the chiral 
string model has been recognized to be an intermediate 
stage between DNG model and that of the generic elastic
model \cite{Carter, Davis1, Carter2}, which has interesting
cosmological implications. The microphysics of this kind
of topological defects has been investigated, opening up
the possibility to have chiral vortons more stable
than vortons  of other kinds \cite{Carter}.

The mathematical generalization of the chiral string model
for an arbitrary number of dimensions is irresistible
and possible which in turn gives a new understanding
of these string features. Recently, the idea of extra 
dimensions has attracted a lot of interest 
because we can think that our universe is like a membrane embedded in 
a higher dimensional spacetime \cite{Randall}. The study
of this ideas is currently under way but, actually,
it is an idea arose from general relativity (GR) a long time ago
which has been pursued by some authors \cite{Regge}, whose
aim was to reformulate GR by means of an embedding of the
spacetime in a higher dimensional space. 

The purpose of this paper 
is extend in an alternative geometrical 
way the dynamical results for chiral strings reported 
in \cite{Carter, Vilenkin} using a Kaluza-Klein (KK)
reduction mechanism \cite{Nielsen} and following
closely the variational techniques developed in 
\cite{Defo, Edges, Defoedges, Supconmem}.
Bearing in mind the KK idea, and assuming our original 
background spacetime to be 4-dimensional, the 
generalization to higher dimensional objects (membranes)
tracing out worldsheets is possible. From this assumption 
we found that the dynamics of the chiral membranes 
resemble that of a five-dimensional DNG case. However, our
analysis can applied to any dimensional fixed background
spacetime and for the membrane with fitting dimension.  
 We describe now our
membrane with an extended embedding. This description
has the advantadge of treat the new membrane on the 
same footing as an ordinary DNG membrane 
\cite{Defo, Defoedges}.
It is often the case that while an existing 
theory admits a number of equivalent descriptions, 
one of them suggets generalizations and simplicities 
more readily than others. This is our goal.

Our membrane described by a higher embedding, exhibits 
a dynamics on the brane caused only by dynamical 
deformations on the membrane itself.
The equations of motion are a generalization of those 
of the DNG type, i.e., the motion of the chiral membranes
looks like minimal surfaces in a KK space but subject 
to a particular condition. In addition, we 
have the current conservation on the membrane which 
emerges as the remain equation of motion for the extra 
variable. Furthermore, to complete the geometric analysis
we shall describe briefly chiral membranes from a 
Hamiltonian point of view taking into account the results 
in \cite{Supconmem}.

The paper is organized as follows. In section 2 we 
develop the essential mathematical features to describe 
the superconducting chiral membranes. This is done by 
exploting the theory of deformations achieved in 
\cite{Defo, Defoedges, Supconmem}. 
In section 3, we present a simplified 
version for the dynamics of chiral extended objects, 
in contrast with others approaches. In section 4 we 
specialize in the chiral string model reproducing 
the content of \cite{Carter, Vilenkin, Davis1} 
showing consistency of our description. Moreover, we found
a new method of integrability for chiral string model using a 
light-cone gauge. In section 5 we found integrability
for a simple chiral domain wall model.
A Hamiltonian approach for chiral membranes 
is developed in section 6. Finally, we give conclusions 
and perspectives of the work.

\vspace{0.4cm}

\section{Geometry for Chiral Membranes}

In this section we describe both the intrinsic and
extrinsic geometry for chiral membranes, i.e.,
possesing null currents on the worldsheet ($\omega =
\gamma^{ab}\, \phi_{,a} \phi_{,b} = 0$), based in the 
Kaluza-Klein approach achieved in \cite{Nielsen}. The
present development is close to the conceptual framework
made in \cite{Defo, Edges, Defoedges, Supconmem}. 
To begin with, we consider a relativistic membrane 
of dimension $d$, whose 
worldsheet $\{ m, \Gamma_{ab} \}$ is an oriented 
timelike $d+1$-dimensional manifold, embedded in a 
5-dimensional extended arbitrary fixed background 
spacetime $\{ {\cal M}, g_{{\bar{\mu} \bar{\nu}}} \}$,
$\bar{\mu}= 0,1,...,4$. We shall describe
the worldsheet by the extended embedding
\begin{equation}
X^{\bar{\mu}}= 
\left(
\begin{array}{l}
X^\mu (\xi^a)\\
\,\,\,\phi (\xi^a)
\end{array}
\right)\,\,,
\label{eq:extembb}
\end{equation}
where $\phi$ is a field living on the worldsheet $m$; 
$a,b= 0,1,2$, and $\xi^a$ are coordinates on the 
worldsheet. With the former 
embedding, we can make contact with the 
KK description for 
the background space-time metric
\begin{equation}
g_{\bar{\mu} \bar{\nu}}= 
\left(
\begin{array}{ll}
g_{\mu \nu}&0 \\
0&g_{44} 
\end{array}
\right)\,\,,
\label{eq:newmetric}
\end{equation}
where $g_{\mu \nu}$ is the metric of the original background 
spacetime, and $g_{44}$ is a constant. 
The tangent basis for the worldsheet is defined by
\begin{equation}
e_a := X^{\bar{\mu}}{}_{,a} \partial_{\bar{\mu}} 
= e^{\bar{\mu}}{}_a \partial_{\bar{\mu}} 
\end{equation}
where the prime denotes a partial derivative with
respect to the coordinates $\xi^a$. The tangent 
vectors $e^{\bar{\mu}}{}_a$, associated with the 
embedding (\ref{eq:extembb}), can be written as
\begin{equation}
e^{\bar{\mu}}{}_a= 
\left(
\begin{array}{l}
e^\mu {}_a\\
\phi_{,a}
\end{array}
\right)\,\,.
\label{eq:extvect}
\end{equation}
The metric induced on $m$ is given by
\begin{equation}
\Gamma_{ab}= g_{\bar{\mu} \bar{\nu}}
e^{\bar{\mu}}{}_a e^{\bar{\nu}}{}_b 
= \gamma_{ab} + g_{44} \phi_{,a}\phi_{,b}\,\,,
\label{eq:newmetric}
\end{equation}
where $\gamma_{ab}= g_{\mu \nu}e^\mu {}_a e^\nu {}_b$
is the standard metric for the worldsheet without
the field $\phi$. 
The normal basis for the worldsheet is denoted
by $n^{\bar{\mu}\,I}$ which is intrinsically 
defined by 
\begin{equation}
g_{\bar{\mu} \bar{\nu}}n^{\bar{\mu}\,I} 
n^{\bar{\nu}\,J}
= \delta^{IJ}\,, \hspace{1cm} g_{\bar{\mu} 
\bar{\nu}}n^{\bar{\mu}\,I} e^{\bar{\nu}}{}_a 
= 0\,,
\end{equation}
where $I,J = 1,2,...,N - d$. We can write explicitly
the complete orthonormal basis, which we label as
$n^{\bar{\mu}\,I} =
\{ n^{\bar{\mu}\,i} , n^{\bar{\mu}\,(4)}\}$ as 
follows,
\begin{equation}
n^{\bar{\mu}\,i}=
\left(
\begin{array}{l}
n^{\mu \,i}\\
0
\end{array}
\right)\,\,,
\hspace{2cm}
n^{\bar{\mu}\,(4)}=
\sqrt{g_{44}} 
\left(
\begin{array}{l}
e^\mu {}_a \phi^{,a} \\
- g^{44}
\end{array}
\right)\,\,,
\label{eq:extvect}
\end{equation}
where we have assumed that $n^{\mu \,i}$ satisfy
$g_{\mu \nu} n^{\mu \,i}n^{\nu \,j}= \delta^{ij}$, and
$i$ take the values $i= 1,...,4-d$.
We write down the corresponding Gauss-Weingarten 
equations for the embedding (\ref{eq:extembb}),
\begin{eqnarray}
D_a e^{\bar{\mu}}{}_b &=& \gamma_{ab} ^c 
e^{\bar{\mu}}{}_c - K_{ab} ^I n^{\bar{\mu}}{}_I \,\,,
\\
D_a n^{\bar{\mu}\,I} &=& K_{ab} ^I e^{\bar{\mu}\,b}
+ \omega_{a}{}^{IJ} n^{\bar{\mu}}{}_J\,\,,
\end{eqnarray}
where $D_a := e^{\bar{\mu}}{}_a D_{\bar{\mu}} $ is 
the gradient along the tangential basis, and 
$ D_{\bar{\mu}} $ is the covariant derivative 
compatible with $g_{\bar{\mu}\bar{\nu}}$, and 
$\gamma_{ab} ^c$ denote the connection coefficients 
compatibles with $\Gamma_{ab}$. 
The extrinsic curvature $K_{ab} ^I$ along the normal 
basis is defined as
\begin{equation}
K_{ab} ^I = - n_{\bar{\mu}} {}^I D_a e^{\bar{\mu}}{}_b \,.
\end{equation}
The last expression can be split as follows: {\it i})
For $I=i$ we have, $K_{ab} ^i= - n_\mu {}^i 
D_a e^{\mu}{}_b$ which is the well known expression
for the extrinsic curvature for the worldsheet of the 
membrane \cite{Defo},
and {\it ii}) for $I=4$, $K_{ab} ^{(4)}
= \sqrt{g_{44}} \,\nabla_a \nabla_b \phi$, where 
$\nabla_a$ is the covariant derivative compatible 
with $\gamma_{ab}$. The index $(4)$ denotes the direction
along the normal $n^{\bar{\mu}\,(4)}$. The extended 
extrinsic twist, 
\begin{equation}
\omega_a {}^{IJ} = g_{\bar{\mu} \bar{\nu}}
n^{\bar{\mu}J}D_a n^{\bar{\nu}I}\,,
\end{equation}
is the connection associated with covariance under
normal rotations. With respect to the last adapted
basis, it is simple to check out that
$\omega_a {}^{ij} = \tilde{\omega}_a {}^{ij}$ (it 
reduces to an extrinsic twist potential in four dimensions) and 
$\omega_a {}^{i\,(4)} = \sqrt{g_{44}}\,\phi^{,\,b}
K_{ab} ^i$, and $\omega_a {}^{(4)(4)} =0$. 
Note that the mixed twist is constructed from
the projection of original worldsheet extrinsic curvature
along the conserved current.

\vspace{0.4cm}

\section{Chiral Membrane Dynamics}

In this section we will show the equivalence between the
chiral membrane dynamics and the DNG dynamics in an 
extended background spacetime plus a chirality condition. 
The starting point to discuss the dynamics of chiral 
membranes is the DNG-like action which is invariant under 
reparametrizations of the worldsheet $m$,
\begin{equation}
S= - \mu_0 \,\int_m d^{d+1} \xi \,\sqrt{- \Gamma} \,\,,
\label{eq:chiaction}
\end{equation}
where $\Gamma$ is the determinant of the induced metric
(\ref{eq:newmetric}) from the space-time by the embedding 
(\ref{eq:extembb}), and $\mu_0$ is a constant.
The determinant is straightforwardly computed and given by
$\Gamma = \gamma \,(1 + g_{44}\,\omega)$, where $\gamma$ 
is the determinant of the old induced metric on the 
worldsheet, $\gamma_{ab}$. The action (\ref{eq:chiaction})
turns into
\begin{equation}
S= - \mu_0 \,\int_m d^{d+1} \xi \sqrt{- \gamma}\,
(1 + g_{44}\,\omega)^{1/2} \,.
\label{eq:cuasi-act}
\end{equation}
Observe that the resulting action from the DNG 
like action (\ref{eq:chiaction}),
is the one for superconducting strings involving 
the Nielsen model, where ${\cal L}(\omega)= 
\sqrt{1 + g_{44}\,\omega}$, \cite{Nielsen}.
In other words, the superconducting string theory
with the Nielsen model is equivalent to DNG like 
action (\ref{eq:chiaction}). 

An important issue that deserves attention is 
that of the equations of motion which are already 
known in \cite{Carter,Vilenkin}. We want rebound 
here the geometrical framework introduced in 
Sec. 2 in the attainment of chiral membrane dynamics.
Using similar variational techniques to that
developed in \cite{Defo,Defoedges} we can get 
the equations of motion from the action 
(\ref{eq:chiaction}). It is worthy to mention that
this method is very graceful because rebound the 
geometrical nature of the worldsheet. The variation
of the action (\ref{eq:chiaction}) gives
\begin{eqnarray}
\delta S &=& -\mu_0 \int_m d^{d+1} \xi \,\frac{1}{2}\,
\sqrt{ -\Gamma}\,\Gamma^{ab} \,\delta \Gamma_{ab} 
\nonumber \\
&=& -\mu_0 \int_m d^{d+1} \xi \,\sqrt{ -\Gamma}\, 
\Gamma^{ab} K_{ab} ^I \Phi_I = 0\,\,,
\end{eqnarray}
where we have considered only normal deformations
to the worldsheet\footnote{The tangential deformations 
can be identified with the actions of worldsheet 
diffeomorphisms  so we can ignore them since we are 
interested in quantities invariant under reparameterizations 
of the worldsheet. These tangential deformations are important 
in the study of composite objects \cite{Defoedges, Cordero4}.}, 
$\Phi^I$ are the deformation
normal vector fields and $\Gamma^{ab}$ is the 
inverse metric of $\Gamma_{ab}$ given by $\Gamma^{ab}
= \gamma^{ab} - g_{44} \nabla^a \phi \nabla^b \phi$. 
We can immediately read the equations of motion
\begin{equation}
\mu_0 \Gamma^{ab} K_{ab} ^I = 0 \,\,.
\label{eq:eqmotion}
\end{equation}
It is worth noticing the similarity of these
equations with those ones arising for minimal surfaces, 
namely, $\gamma^{ab} K_{ab} ^i = K^i = 0$, \cite{Defo}.
In fact, in our description $\Gamma^{ab}$ play 
the role of a metric. Let us now decode the 
important cases involved in the Eq. (\ref{eq:eqmotion}).
a) $I=i$. The equations of motion take the form,
\begin{equation}
\mu_0 \gamma^{ab} K_{ab} ^i - \mu_0 g_{44}\nabla^a \phi 
\nabla^b \phi K_{ab} ^i = 0\,.
\label{eq:eqmotion1}
\end{equation}
On other hand, in the generic superconducting 
membranes picture, the strees-energy-momentum tensor 
adquires the form $T_{ab} = {\cal L}(\omega) \gamma_{ab} 
- 2 \,(d {\cal L}/d \omega)\, \nabla_a \phi 
\nabla_b \phi$, where ${\cal L}(\omega)$ is a function
of $\omega$, depending on the 
particular models \cite{Carter2, Cordero}. 
When the chiral current limit is taken into account, 
the quantities ${\cal L}(\omega)$ and 
${d {\cal L}}/{d \omega}$, adquire constant 
values. If we define 
$g_{44}:= 2(d {\cal L}/d \omega)|_{\omega =0}$
and $\mu_0 = {\cal L}(\omega)|_{\omega =0}$, we can identify 
the Eq. (\ref{eq:eqmotion1})
with the standard equations of motion, namely:
${ T}^{ab} K_{ab} ^i = 0$, \cite{Carter2,Cordero}.
b) $I=(4)$. In this case we have now directly,
\begin{equation}
\Gamma^{ab} K_{ab} ^{(4)} =0= \nabla_a \nabla^a
\phi \,,
\end{equation}
which is a wave equation for $\phi$, corresponding
to a conserved current carrying onto the 
worldsheet for chiral currents.

\section{Chiral String Model}

With the purpose of make contact with previous works 
\cite{Carter, Vilenkin}, we specialize now to the case 
of chiral strings. In the next section we will study
the case of chiral domain walls. We illustrate the 
chiral string model from a Lagrangian point of view. 

\subsection{Gauge Choices}

The presence of the gauge symmetry in a field theory
means that not all of the field components, $X^{\bar{\mu}}
(\xi^a)$, are dynamical. In our case, the 
reparameterization invariance allow us to choose 
a gauge in which the dynamical equations are 
tractable. Due to we have considered a DNG like action, 
(\ref{eq:chiaction}), we have the freedom of choice of
an acceptable gauge condition. 

{\bf{a)}} Recently, Carter and Peter presented 
a solution for the chiral string model. This is 
an ansatz useful in the solution for the transonic string 
model \cite{Carter}. We reproduce the same result
via a particular choice for the embedding  of the
worldsheet in the extended background. Assuming the
embedding
\begin{equation}
X^{\bar{\mu}}= 
\left(
\begin{array}{l}
X^\mu (q,\eta)\\
\,\,\,\,\,\, \alpha \eta
\end{array}
\right)\,\,,
\label{eq:nextembb}
\end{equation}
where $q$ and $\eta$ are coordinates on the 
worldsheet and $\alpha$ is a constant of
proportionality. With this choice, the internal 
scalar field is promoted as $\phi = \alpha \eta$.
In such a case, and according to our notation,
we demand the condition
\begin{equation}
\sqrt{-\Gamma}\, \Gamma^{ab}= 
\left(
\begin{array}{ll}
0&  1 \\
1 &0 
\end{array}
\right)\,\,.
\label{eq:ngaugeinvgamma}
\end{equation}
The induced metric $\gamma_{ab}$, will take
the form
\begin{equation}
\gamma_{ab}= 
\left(
\begin{array}{ll}
\,\,\,0&  \sqrt{- \Gamma} \\
\sqrt{- \Gamma} &- g_{44}
\end{array}
\right)\,\,,
\end{equation}
with the inverse matrix given by
\begin{equation}
\gamma \,\gamma^{ab}= 
\left(
\begin{array}{ll}
\,\,- g_{44}& - \sqrt{- \Gamma} \\
-\sqrt{- \Gamma} &\hspace{.4cm}\,\,0
\end{array}
\right)\,\,.
\end{equation}
Using the last information we observe easily the 
condition for chirality to be
\begin{equation}
\gamma^{\eta \eta} = 0\,.
\label{eq:newcondit}
\end{equation}
The corresponding equations of motion for
(\ref{eq:nextembb}) taking into account the gauge
(\ref{eq:ngaugeinvgamma}), result in
\begin{equation}
\partial_q \partial_\eta \,X^{\bar{\mu}} = 0\,,
\end{equation}
which have been recently reported in 
\cite{Carter}.

{\bf{b)}} In a short time later, Blanco-Pillado et al.,
\cite{Vilenkin}, gave an alternative description and a 
solution for the same model by means of a different 
gauge choice of that proposed in \cite{Carter}. In our 
description, a convenient gauge is the well known 
conformal gauge,
\begin{equation} 
\sqrt{- \Gamma}\, \Gamma^{ab}= \eta^{ab}\,\,.
\end{equation}
Following the standard string theory 
notation\footnote{As is well known, in such a 
case the worldsheet is parametrized by the coordinates
$\xi^0 = \tau$ and $\xi^1 = \sigma$. The symbols $^.$ and
$'$ denote partial derivatives with respect to $\xi^0$
and $\xi^1$, respectively.}, we attempt to solve
the chiral string model from (\ref{eq:chiaction}), 
which is accomplished taking into account the next 
form for the metric
\begin{equation}
\gamma_{ab}= 
\left(
\begin{array}{ll}
-\sqrt{-\Gamma} - g_{44}\,{\phi '}^2 &  
\,\,\,\,\,\,\,\,-g_{44}\,\dot{\phi}\,\phi ' \\
\,\,\,\,\,\,\,\,\,- g_{44}\,\dot{\phi}\,\phi '  &
\sqrt{-\Gamma} - g_{44}\,\dot{\phi}^2 
\end{array}
\right)\,\,,
\label{eq:gaugegamma}
\end{equation}
{i.e.,} this corresponds to our gauge choice.
In fact, this choice is equivalent to that 
studied in \cite{Vilenkin}. 
To complement the calculation we write the 
inverse metric $\gamma^{ab}$, easily obtained,
\begin{equation}
\gamma \gamma^{ab}= 
\left(
\begin{array}{ll}
\sqrt{-\Gamma} - g_{44}\,\dot{\phi}^2 &  
\,\,\,\,\,\,\,\,\,\,\,\,g_{44}\,\dot{\phi}\,\phi ' \\
\,\,\,\,\,\,\,\,\,\,g_{44}\,\dot{\phi}\,\phi '  &
-\sqrt{-\Gamma} - g_{44}\,{\phi '}^2 
\end{array}
\right)\,\,.
\label{eq:gaugeinvgamma}
\end{equation}
Taking into account (\ref{eq:gaugeinvgamma}), the 
condition for chirality $(\omega=0)$ will take the 
form
\begin{equation}
\dot{\phi}^2 - {\phi'} ^2 = 0\,,
\label{eq:wavephi}
\end{equation}
which is a condition for the field $\phi$ in 
this gauge. The equations of motion for the 
system, using the refered gauge, are obtained in 
a straightforward
way from those of DNG case, $\partial_a ( \sqrt{- \Gamma}
\Gamma^{ab}\,X^{\bar{\mu}}{}_b ) = 0$, namely, 
$\ddot{X}^{\bar{\mu}} - {X} ^{''\bar{\mu}} = 0$\,, 
whose solutions are $X^{\bar{\mu}} = \frac{1}{2}a^{\bar{\mu}}
(\tau + \sigma) + \frac{1}{2}b^{\bar{\mu}}(\tau - \sigma)$. 
The conformal gauge imposes the following constraints 
over $a^{\bar{\mu}}$ and $b^{\bar{\mu}}$, namely,
$a^{\bar{\mu}}`a`_{\bar{\mu}}= 1$ and 
$b^{\bar{\mu}}`b`_{\bar{\mu}}= 1$, where ` denotes 
derivation with respect to their arguments. These are the 
same conditions like in the DNG case but now in the 
extended space. Explicitly they take the form 
$|\vec{a} `| ^2 + g_{44} \phi_1 ^{`\,\,2} = 1$ and 
$|\vec{b} `| ^2 + g_{44} \phi_2 ^{`\,\,2} = 1 $,  
where $\phi_1 $ and $\phi_2$ are the last components
of the vectors, $a^{\bar{\mu}}= ( \vec{a} \,,\,\phi_1)$ 
and $b^{\bar{\mu}}= ( \vec{b} \,,\,\phi_2)$. However, 
the condition (\ref{eq:wavephi}) tell us that either $\phi_1
(\tau + \sigma) =0$ or $\phi_2(\tau - \sigma) =0$, and 
we are able to reproduce the results 
of \cite{Vilenkin}. The vorton states are obtained when 
$\dot{X} ^\mu = X`^{\mu} $ i.e., $\vec{b}` =0$ 
and $\phi_1 =0$.

{\bf{c)}} Now we turn to consider the light-cone 
gauge over the spacetime coordinates adapted to 
our description. 
When we study non-superconducting strings, it is 
well known that the orthonormal gauge do not 
fully fix the gauge because there is residual 
reparametrization invariance. A favorite gauge 
choice that fix the gauge and allow us to solve 
the constraints is the light-cone gauge 
\cite{GSW}. In string theory, this gauge is 
convenient for its quantization because it allow 
us eliminate all unphysical degreees of freedom 
and unitarity is guaranteed. In other context, 
this gauge was used by Hoppe in the search for
explicit solutions for the classical equations
of motion of relativistic membranes \cite{Hoppe}.
We shall use this orthonormal light-cone gauge
in the search of integrability for the chiral
string model.

To proceed further,
we assume the original background metric to be flat,
$g_{\mu \nu} = \eta_{\mu \nu}$, with signature
$(-,+,+,+)$ and the embedding (\ref{eq:extembb}).
For the KK spacetime we define light-cone 
coordinates, $X^+$ and $X^-$, as
\begin{eqnarray}
X^+ &=& \frac{1}{\sqrt{2}}\,(X^0 + 
       \sqrt{g_{44}}\,\phi) \,, 
\label{eq:X+}       \\
X^- &=& \frac{1}{\sqrt{2}}\,(X^0 - 
       \sqrt{g_{44}}\,\phi)\,,
\label{eq:X-}      \\      
\vec{X} &=& (X^1,X^2,X^3)\,. 
\end{eqnarray}               
The light-cone gauge points $\tau$ along $X^+$,
\begin{equation}
X^+ = X^+ _0 + P^+ \,\tau \,,
\end{equation}
where $X^+ _0$ and $P^+$ are constants. The 
idea is solve for $X^-$ leaving the $X^i$
variables, where $i=1,2,3$. Laying hold of 
the orthonormal light-cone gauge \cite{Hoppe},
\begin{equation}
\Gamma_{ab}= 
\left(
\begin{array}{ll}
\Gamma_{\tau \tau} & \,\, 0\\
\,0  & \Gamma_{AB}
\end{array}
\right)\,\,,
\label{eq:algo}
\end{equation}
where $\Gamma_{ab}$ is given by 
(\ref{eq:newmetric})
and $A,B=1,...,d$, besides of 
$\sqrt{-\Gamma}\,\Gamma^{\tau \tau} = -1$, we can simplify 
the equations of motion, $\partial_a(\sqrt{-\Gamma}\,
\Gamma^{ab}\,X^{\bar{\mu}}{} _{,b})= 0$ in the set of 
equations
\begin{eqnarray}
{\cal D} \,X^{\bar{\mu}} &=& 0 \,,
\label{eq:lcemotion} 
\\
2 P^+ \dot{X} ^- &=& \dot{\vec{X}}\cdot \dot{\vec{X}}
                    + \bar{\Gamma} \,,
\label{eq:constraint1}                    
\\
P^+ X^- {}_{,A}&=& \dot{\vec{X}}\cdot \vec{X}_{,A}
\label{eq:constraint2}
\end{eqnarray}
where we have defined $\bar{\Gamma}:= {\mbox{det}}
(\Gamma_{AB}) = - \Gamma_{\tau \tau}$ and we have 
defined the differential operator
\begin{equation}
{\cal D} := - \partial_\tau ^2 + \partial_A 
(\bar{\Gamma}\,\Gamma^{AB}\,\partial_B )\,.
\end{equation}
 Equation (\ref{eq:lcemotion}) represents the 
equations of motion in this gauge, 
(\ref{eq:constraint1}) and (\ref{eq:constraint2})
are the constraints relations for the system.
Deriving with respect to $\tau$ the Eq. 
(\ref{eq:constraint1}), we can rewrite it as
$P^+ {\cal D} X^- = \dot{\vec{X}}\cdot {\cal D}
\vec{X}$; so if
\begin{equation}
{\cal D} \vec{X} =0 \,,
\label{eq:lcemotion1}
\end{equation}
we get the condition ${\cal D} X^- = 0$, 
which we can observe from (\ref{eq:lcemotion}).
Thus, we have reduced the problem to solve the set 
(\ref{eq:constraint1}), (\ref{eq:constraint2}) 
and (\ref{eq:lcemotion1}). So far the results 
are general for minimal surfaces of arbitrary
dimension. Now we specialize to the case of 
chiral strings. In order to get integrability
for the chiral string model, besides Eqs. 
(\ref{eq:constraint1}), (\ref{eq:constraint2}) 
and (\ref{eq:lcemotion1}), the condition $\omega=0$
must be considered. Using the stringy notation,
the appropiate expressions are
\begin{equation}
\Gamma_{ab}= 
\left(
\begin{array}{ll}
- \vec{X}' \cdot \vec{X}' & \,\,\,\,\,\,\,\,0\\
\,\,\,\,\,\,\,\,0  & \vec{X}' \cdot \vec{X}'
\end{array}
\right)\,\,,
\label{eq:algomas}
\end{equation}
with $\Gamma_{\tau \tau}= - \Gamma_{\sigma \sigma}$.
The equations of motion (\ref{eq:lcemotion1})
transform in the wave equation: ${\cal D} \vec{X}
= (- \partial_\tau ^2 + \partial_\sigma ^2 )\vec{X}
= -\ddot{\vec{X}} +\vec{X}'' =0$, whose general 
solution is given by
\begin{equation}
\vec{X} = \vec{a} (\tau + \sigma) + \vec{b}(\tau -
\sigma) \,.
\end{equation}
The constraints (\ref{eq:constraint1}) and
(\ref{eq:constraint2}) adquire the form
\begin{eqnarray}
P^+ \dot{X}^- &=& |\vec{a}`|^2 + |\vec{b}`|^2 \,,
\label{eq:constraint3} \\
P^+  X^{- '}  &=& |\vec{a}`|^2 - |\vec{b}`|^2 \,,
\label{eq:constraint4}
\end{eqnarray}
where we have used the notation $\vec{a}`$ and 
$\vec{b}`$ to denote derivatives with respect to 
their arguments. From the Eqs. (\ref{eq:X+}) and
(\ref{eq:X-}) as well as Eq. (\ref{eq:newmetric}),
we can separate the metric $\gamma_{ab}$,
\begin{equation}
\gamma_{ab}= 
\left(
\begin{array}{ll}
- \vec{X}' \cdot \vec{X}' - \frac{1}{2}
\,(P^+ - \dot{X} ^-)^2 & \,\,\,\frac{1}{2}
(P^+ - \dot{X} ^-) X^{- '}\\
\,\,\,\,\,\,\,\,\,\,\,\,\frac{1}{2}\,(P^+ - 
\dot{X} ^-) X^{- '}  & 
\vec{X}' \cdot \vec{X}' - \frac{1}{2}\,(X^{-'})^2
\end{array}
\right)\,\,,
\label{eq:algomas1}
\end{equation}
and its inverse
\begin{equation}
\gamma \gamma^{ab}= 
\left(
\begin{array}{ll}
\vec{X}' \cdot \vec{X}' - \frac{1}{2}\,(X^{-'})^2 & 
\,\,\,\,\,\,\,\,-\frac{1}{2}\,(P^+ - 
\dot{X} ^-) X^{- '}\\
-\frac{1}{2}\,(P^+ - \dot{X} ^-) X^{- '}  & 
- \vec{X}' \cdot \vec{X}' - \frac{1}{2}
\,(P^+ - \dot{X} ^-)^2 
\end{array}
\right)\,\,.
\label{eq:algomas2}
\end{equation}
Now, the chirality condition becomes
\begin{eqnarray}
\omega &=& \gamma^{ab}\,\nabla_a \phi
\nabla_b \phi \nonumber \\
&=& \frac{1}{2\gamma g_{44}}\,[
(P^+ - \dot{\vec{X}}^2 )^2 - (X^{-'})^2]
(\vec{X}' \cdot \vec{X}') =0 \,,
\end{eqnarray}
from which is deduced that
\begin{equation}
X^{-'}= \pm (P^+ - \dot{\vec{X}}^2)\,.
\label{eq:solution}
\end{equation}
Plugging the Eq. (\ref{eq:solution}) in the constraint
(\ref{eq:constraint2}) we get the conditions that should
satisfy $\vec{a}$ and $\vec{b}$ in the chiral string
solution, namely,
\begin{equation}
\pm (P^+)^2 = (\vec{a}` + \vec{b}`)\cdot [ \vec{a}` 
- \vec{b}` \pm 
P^+ (\vec{a}` + \vec{b}`)]\,.
\label{eq:condition}
\end{equation}
Thus, this equation suggest to consider some cases. 

For instance, if we assume $P^+=1$, 
we get
\begin{equation}
1= 2(\vec{a}` + \vec{b}`)\cdot \vec{a}`  \,,
\end{equation}
which is different for the values for $\vec{a}$ and 
$\vec{b}$ reported in \cite{Vilenkin}. 

It is worthy to mention that the integrability
for the last cases was reached in absence of
electromagnetic field coupled to superconducting
strings. From the equations defining $X^+$, $X^-$ and the 
contraints (\ref{eq:constraint3}) and 
(\ref{eq:constraint4}) we can find the value for $\phi$. 
Again the vortons states are obtained when $\vec{b} `=0$ 
and  in such case $\vec{a}$ satisfy the relation 
$\pm P^{+2} = \vec{a}`\cdot [\vec{a} ` ( 1 \pm P^+)] $.

\section{Chiral Domain Wall Model}

In this section applying a similar mechanism
to that for conformal gauge chiral string model 
we get integrability for a chiral domain wall
model by means of a special ansatz. For
such an intention, with all the former ingredients, 
we study a worldsheet described by the embedding
\cite{Vilenkin1},
\begin{equation}
X^{\bar{\mu}}(\tau,\xi^1,\xi^2)= 
\left(
\begin{array}{l}
\,\,\tau\\
\vec{\tilde{X}}
\end{array}
\right)\,\,,
\label{eq:dwembedd1}
\end{equation}
where
\begin{equation}
\vec{\tilde{X}}= 
\left(
\begin{array}{l}
\vec{{X}} \\
\phi
\end{array}
\right)\,.
\label{eq:dwembedd2}
\end{equation}
Now we can choose a gauge similar to the 
conformal gauge for strings , as follows 
\begin{equation}
\Gamma_{ab}= 
\left(
\begin{array}{ll}
\Gamma_{\tau \tau} & \,\, 0\\
\,0  & \Gamma_{AB}
\end{array}
\right)\,\,,
\label{eq:metric1}
\end{equation}
where $\Gamma_{AB} = g_{\bar{\mu} \bar{\nu}} 
X^{\bar{\mu}}{} _{,A}X^{\bar{\nu}}{}_{,B}$
and $\Gamma_{\tau A} =0$. We assume the special 
form for the embedding (\ref{eq:dwembedd1}),
as 
\begin{equation}
\vec{\tilde{X}}= \hat{\tilde{n}}\,\xi^2 + 
\vec{\tilde{X}}_{\perp} (\tau,\xi^1) \,,
\label{eq:dwgauge}
\end{equation}
with the conditions on $\hat{\tilde{n}}$ to be
a unit vector and perpendicular to 
$\vec{\tilde{X}}_{\perp}$. So, in this conformal 
gauge we have the constraints
\beq
\dot{\vec{\tilde{X}}}_{\perp} \cdot 
{\vec{\tilde{X}}} '_{\perp} &=& 0 \,,
\label{eq:const1} \\
\dot{\vec{\tilde{X}}}_{\perp} \cdot 
\dot{{\vec{\tilde{X}}}}_{\perp} + 
{\vec{\tilde{X}}}'_{\perp} \cdot 
{\vec{\tilde{X}}} '_{\perp} &=& 1 \,,
\label{eq:const2}
\eeq
and the condition $\sqrt{- \Gamma}\,\Gamma^{\tau \tau} 
= - 1$. It is straightforward to demonstrate that 
$\bar{\Gamma}:= {\mbox{det}}(\Gamma_{AB}) = - 
\Gamma_{\tau \tau}$. In this part of the work, $^.$ and
$'$ denote derivatives with respect to $\tau$ and $\xi^1$,
respectively.

Assuming $g_{\mu \nu}$ to be Minkowski's metric and resting in the
constraints (\ref{eq:const1}) and (\ref{eq:const2}), the
induced metric $\Gamma_{ab}$ takes the form,
\begin{equation}
\Gamma_{ab}= 
\left(
\begin{array}{lll}
-|\vec{\tilde{X}} _\perp '|^2 & \hspace{4mm} 0 & 0\\
\hspace{6mm} 0  & |\vec{\tilde{X}} _\perp '|^2 & 0 \\
\hspace{6mm} 0 & \hspace{4mm} 0 & 1
\end{array}
\right)\,.
\label{eq:dwmetric}
\end{equation}
According to the standard DNG equations of motion, in
our present case the corresponding ones are 
promoted as $\ddot{\vec{\tilde{X}}} _\perp  - 
\vec{\tilde{X}} '' _\perp =0$,
whose solutions have the form 
\begin{equation}
\vec{\tilde{X}} _\perp = \frac{1}{2}\vec{\tilde{a}} 
( t + \xi^1) + \frac{1}{2}\vec{\tilde{b}} 
( t - \xi^1) \,.
\label{eq:dwsolution}
\end{equation}
Imposition of the chirality for superconducting domain
walls, lead us to the relation
\be
|\vec{\tilde{X}}_\perp {}'|^2 (n^4)^2 = 
\dot{\phi}_\perp ^2 - {\phi}''_\perp {}^2 \,,
\label{eq:dwcondition}
\ee
where $n^4$ is the four component of the vector
$\hat{\tilde{n}}$. Furthermore, the constraints 
(\ref{eq:const1}) and 
(\ref{eq:const2})  read as $ |\vec{\tilde{a}}{}{}'|^2 
=1$ and $ |\vec{\tilde{b}}{}{}'|^2 =1$, or explicitly 
they are given by
\begin{eqnarray}
|\vec{a}{}'|^2  + g_{44} \phi ' {}^2 &=& 1\,, \\
|\vec{{b}}{}'|^2  + g_{44} \bar{\phi} ' {}^2 &=&1 \,,
\end{eqnarray}
where we have considered the notation $\vec{\tilde{a}} = 
(\vec{a} ,\,\, \phi)$ and $\vec{\tilde{b}} = 
(\vec{b} ,\,\, \bar{\phi})$, {\it i.e.,}
\begin{equation}
\vec{\tilde{X}}_\perp= 
\left(
\begin{array}{l}
\vec{{X}}_\perp = \frac{1}{2}\vec{{a}} ( t + \xi^1) + 
\frac{1}{2}\vec{b} ( t - \xi^1)
\\
\phi_\perp = \frac{1}{2}\phi( t + \xi^1) + 
\frac{1}{2}\bar{\phi}( t - \xi^1)
\end{array}
\right)\,.
\label{eq:dwvect1}
\end{equation}
Plugging (\ref{eq:dwvect1}) in the condition 
(\ref{eq:dwcondition}), the chirality condition is
expressed now as 
\be
\phi ` \,\bar{\phi} ` [1 + g_{44} (n^4)^2] = (n^4) ^2 
(1- \vec{a} ` \cdot \vec{b}` ) \,.
\ee
In a similar way as in the chiral string model case,
the last equation suggests some cases. For example,
the solution considering $n^4 =0$ and $\phi =0$ 
(or $\bar{\phi} =0$), correspond to a straight 
superconducting domain wall with a carrying current 
arbitrary cross-section, but not including current 
along the $\xi^2$ direction.

\section{Hamiltonian Dynamics for Chiral \\
Membranes}

In this part of the work we want to make a Hamiltonian
approach to the dynamics for chiral membranes. 
Nowadays, canonical formulation is successfully
used as a starting point for quantization but, also is used
as a tool for tackling dynamical problems. Bearing in mind
the last use as well as its possible aplication to brane universe scenario,
we first review briefly the Hamiltonian framework
already  developed in \cite{Supconmem}.
Before going to the canonical analysis we shall mention something
about the useful mathematical issues necessary for this purpose. Taking 
into account the established notation of Sect. 2, we will 
consider a superconducting relativistic membrane of 
dimension $d$ whose
worldsheet $\{ m',\gamma_{ab} \}$ is $d+1$-dimensional
which is embedded in a given fixed background spacetime 
$\{ {\cal{M}}',g_{\mu ,\nu} \}$. According to the ADM procedure
for the canonical general relativity, we assume that $m'$ has an
adequate topology such that we can foliate to the worldsheet
into $d$-dimensional spacelike hypersurfaces $\Sigma_t$, 
parametetrized by constant values of a function $t$. Each slice
of the foliation represents the system at an instant of time
and each one is diffeomorphic to each other. In order to describe 
the evolution of the leaves of the foliation is convenient
to introduce the worldsheet time vector field
\be
\label{eq:timevect}
t^a = \dot{X}^a:= N \eta^a + N^A \epsilon^a {}_A \,,
\ee
where $N$ and $N^A$ denote the well-known lapse function
and shift vector, respectively, $\eta^a $ denotes the unit
future-oriented timelike normal vector field to the slice
$\Sigma_t$ and $\epsilon^a {}_A$ the corresponding tangent 
vector.

The decomposition of the several 
geometric quantities involved in the theory in the normal and
tangential parts to the slice, taking advantadge
of the deformation vector field (\ref{eq:timevect}), is
primordial in the Hamiltonian treatment. The various
geometrical quantities that characterize the intrinsic and
extrinsic geometry of $\Sigma_t$, can be decomposed based 
in the formalism of deformations. For instance, the worldsheet 
metric $\gamma_{ab}$ in the ADM decomposition looks like
\be
\gamma_{ab}=\left(
\begin{array}{ll}   
 - N^2 + N^A N_A & N_A \\
\,\,\,\,\,\,\,\,\,\,\,\,\,\, N_A & h_{AB}
\end{array}
\right) \,, 
\label{eq:ggamma}
\ee
and for the inverse
\begin{equation}
\gamma^{ab} =
{1 \over N^2} \left(
\begin{array}{ll}
 - 1  & \hspace{1.5cm}N^A  \\
 N^A   & (h^{AB} N^2 - N^A N^B )
\end{array}
\right)\,.   
\end{equation}
Note that from~(\ref{eq:ggamma}) it follows that:
$\gamma = - N^2 \,h,$
where $h$ is the 
determinant of the metric $h_{AB}$ induced on the 
hypersurface $\Sigma_t$, via the embedding $x^\mu=
X^\mu(u^A)$, and we have assumed that $u^A$ are local
coordinates on $\Sigma_t$. 
For a general treatment
concerning this formalism, see the Refs. \cite{Supconmem,
ADM}. As the starting point for our Hamiltonian
study we have the generic action 
\begin{equation}
S = \int_{m'} d^{d+1}\xi \,\sqrt{- \gamma}\,{\cal L}(\omega) \,,
\label{eq:genericact}
\end{equation}
where ${\cal L}(\omega)$ is a function depending of the
internal and external fields acting on the worldsheet. The
split action (\ref{eq:genericact}) in time and space
is performed according to the geometric canonical 
procedure. The Lagrangian density in the ADM fashion 
is given by
\begin{equation}
L= N\sqrt{h}\,{\cal L}(\omega) \,,
\end{equation}
where we should understand that $ {\cal L}
(\omega)$ has been split in space and time
where $\omega$ has the expression:
$\omega = - (1/N^2) \,[ \dot{\phi} - N^A
\bar{{\cal D}}_A \phi + (t\cdot A) ]^2 +
\tilde{\omega}$, where
$\bar{\omega} = h^{AB} \phi_{,A}\phi_{,B}$.

The phase space ${\mathcal{A}}= \{ X,P;\phi,\pi \}$ is closely
related to the geometry of $\Sigma_t$, which is seen from the  
equations defining the momenta $\pi =- 2 \sqrt{h} 
(\bar{\omega} - \omega )^{1/2}
\,d {\cal L}/d \omega $ and $P_\mu = 
[ - \sqrt{h} {\cal L}(\omega) + (\bar{\omega} -
\omega )^{1/2}\,\pi ]\,\eta_\mu - \pi \,
\tilde{{\cal D}}_A \phi \,\,
\epsilon_\mu {}^A  + \,\,\,\pi  A_\mu $,
conjugate to $\phi$ and $X^\mu$, respectively.

The canonical Hamiltonian vanishes identically which 
stems from the requirement of reparametrization invariance
of the worldsheet. 
For this field theory, the constraints
are given by
\begin{eqnarray}
{\cal C}_0 &=& g^{\mu \, \nu}\Upsilon_\mu \Upsilon_\nu + 
h\,\left( {\cal L}(\omega ) + \frac{\pi ^2}{2\,h\,
(d{\cal L}/d\omega )}
\right)^2  -  \pi^2 \,\left( \omega + \frac{\pi^2}
{4\,h\,(d{\cal L}/d\omega)^2}
\right) 
\label{C0} \\
{\cal C}_A&=& \Upsilon_\mu \,\epsilon^\mu {}_A + \pi\,
\tilde{{\cal D}}_A \phi \,,
\label{C1} 
\end{eqnarray}
where the kinetic momentum is $\Upsilon_\mu := 
P_\mu - \pi A_\mu $, with $A_\mu$ being a background 
electromagnetic potential.

We turn now to consider the canonical description 
for chiral membranes except the background 
electromagnetic field, from two different alternatives, 
in like manner to that for canonical DNG case. 
First, we follow the standard way close to that 
described previously. Next, we benefit from the 
canonical approach developed at the beggining of this 
section using the 
constraints (\ref{C0}-\ref{C1}) directly.
For the Hamiltonian approach for the action 
(\ref{eq:chiaction}), it would be first
decomposed in the ADM fashion as
\be
S= \int_R \int_{\Sigma_t} {\cal N} \sqrt{{\cal H}} \,,
\label{eq:chiralADM}
\ee
where we have defined a new lapse function 
${\cal N}:= N \,(1 + g_{44}\,\omega)^{1/2}/(1 + g_{44}\,
\bar{\omega})^{1/2}$ and ${\cal H}$ is the determinant 
of the induced metric on the hypersurface, 
${\cal H}_{AB}$, from the embedding (\ref{eq:extembb}). 
Explicitly, it is given by 
${\cal H} = h\,(1 + g_{44}\,\bar{\omega})$.
The next step is the computation of the momenta conjugate 
to the configuration variables. They are given by
\begin{eqnarray}
\pi &:=& -  \frac{\sqrt{\cal{H}}\,g_{44}\,(\bar{\omega} 
- \omega)^{1/2}}{(1 + g_{44}\,\bar{\omega})^{1/2}
(1 + g_{44}\,\omega)^{1/2}} 
\label{eq:chiralpi}
\\
P_\mu &:=& \left[ - \sqrt{\cal{H}}\,\frac{(1 + g_{44}\,\omega)^{1/2}}
{(1 + g_{44}\,\bar{\omega})^{1/2}} + \pi \,(\bar{\omega} -
\omega)^{1/2} \right] \,\eta_\mu - \pi {\cal{D}}_A \phi \, 
\epsilon_\mu {}^A\,.
\label{eq:chiralP}
\end{eqnarray}
Taking into account the equations defining the momenta 
(\ref{eq:chiralpi}-\ref{eq:chiralP}) and the chiral
limit ($\omega =0$), we get the constraints
\begin{eqnarray}
C_0 &=& g^{\bar{\mu} \, \bar{\nu}} P_{\bar{\mu}} 
P_{\bar{\nu}} + {\cal H}\,\,,
\label{eq:newC0}
\\
C_A  &:=& P_{\bar{\mu}} \epsilon^{\bar{\mu}}{}_A \,\,,
\label{eq:newCA}
\end{eqnarray}
where we have defined the extended momenta, 
$P_{\bar{\mu}} := (P_\mu , \pi)$.

The second way to get the constraints for the chiral
case is straightforward 
from the constraints (\ref{C0}-\ref{C1}) governing the 
superconducting membrane theory because those encode the 
geometrical information for such. For chiral membranes, 
it is enough to make $\omega =0$ in the expressions 
for ${\cal L}(\omega)$ and $d {\cal L} /\omega $ which 
become constants $c_1$ and $c_2$,
respectively, whose inclusion in the constraints 
provide the corresponding constraints for chiral 
membranes,
\begin{eqnarray}
C_0 = {\cal C}_0 \mid _{\omega =0} &=&
g^{\mu \, \nu}P_\mu P_\nu + h\,c_1 ^2 + 
\frac{c_1}{c_2}\,\pi^2 
:= g^{\bar{\mu} \, \bar{\nu}} P_{\bar{\mu}} 
P_{\bar{\nu}} + {\cal H}\, c_1 ^2 \,\,,
\label{eq:chiC0}
\\
C_A = {\cal C}_A \mid _{\omega =0} &:=& P_
{\bar{\mu}} \epsilon^{\bar{\mu}}{}_A \,\,,
\label{eq:chiC1}
\end{eqnarray}
where the expression
defining the momentum $\pi$, namely, $\pi= -2
\sqrt{h} \,({\bar{\omega}})^{1/2} \, 
(d{\cal L}/d\omega)$ has been used, and we 
have defined $g^{44}$ 
as the quotient $c_1 /2c_2$. Note the similarity
of (\ref{eq:chiC0}) and (\ref{eq:chiC1}) with the 
DNG constraints in the canonical language
where $c_1$ plays the role of tension of the membrane
\cite{ADM}.
The preceding description 
show us that the dynamical behaviour for chiral 
membranes is similar to DNG case,
but their internal structures are different. In fact,
for the chiral string model it is an intermediate stage
between the transonic string model \cite{Carter1} (possesing
two internal degrees of freedom) and that of DNG case
(without internal degrees of freedom).

\section{Conclusions}

In this paper we have developed the dynamics of 
chiral membranes using geometrical techniques.
We are able to reproduce the results of \cite{Carter,Vilenkin, 
Davis1} showing consistency of our description. In 
fact, our scheme is resemble to DNG theory in five 
dimensions using a Kaluza-Klein approach but different 
internal structure. However, our 
study is valid in any number of dimensions for the 
background spacetime and for the embedded membrane.
Integrability for both chiral string model and 
chiral domain wall model
was obtained using a simple ansatz.
The full physical description is not over yet because 
a deep understanding of integration of equations 
of motion has not been accompplished. Finally, we remark that 
our study might be useful in the 
study of brane universe scenarios, at least at canonical level,
 where we could think our world embedded in a 
higher-dimensional space, where the Hamiltonian analysis
is demanded for quantization. Besides this, the search of new 
solutions for the case of chiral superconducting membranes 
is part of a forthcoming paper.

\section{Acknowledgements}

The authors are indebted to R. Capovilla, X. Martin and 
J.J. Blanco-Pillado for many valuable discussions and 
suggestions. E. Rojas 
expresses grateful thanks to Professors A. P. Balachandran and 
Alberto Garc\'{\i}a for the encouragement to the paper 
likewise to the Department of 
Physics of Syracuse University for hospitality. R.C. 
thanks to J. Mendoza and R. Rodriguez for stimulating 
discussions. The support in part from SNI (M\'exico) and 
CONACYT is grateful.

\end{document}